\documentclass[3p]{elsarticle}
\usepackage{url}
\usepackage{graphicx}
\usepackage{hyperref}

\journal{Journal of Proteomics}

\begin{document}

\begin{frontmatter}
\title{An Optimized Data Structure for High Throughput 3D Proteomics Data:
mzRTree}
\author{Sara Nasso\corref{cor1}}
\author{Francesco Silvestri}
\author{Francesco Tisiot}
\author{Barbara Di Camillo}
\author{Andrea Pietracaprina}
\author{Gianna Maria Toffolo}
\address{Department of Information Engineering\\ University of Padova\\
Via Gradenigo 6/B, 35131  Padova Italy}
\cortext[cor1]{Corresponding author:  via Gradenigo 6/B,
35131  Padova Italy; +39-049-8277834 (voice); +39-049-8277826 (fax);
\url{sara.nasso@dei.unipd.it}.}

\begin{abstract}
As an emerging field, MS-based proteomics still requires software tools for
efficiently storing and accessing experimental data. In this work, we focus on
the management of LC-MS data, which are typically made available in standard
XML-based portable formats. The structures that are currently employed to manage
these data can be highly inefficient, especially when dealing with
high-throughput profile data. LC-MS datasets are usually accessed through 2D
range queries. Optimizing this type of operation could dramatically reduce the
complexity of data analysis. We propose a novel data structure for LC-MS
datasets, called mzRTree, which embodies a scalable index based on the R-tree
data structure. mzRTree can be efficiently created from the XML-based data
formats and it is suitable for handling very large datasets. We experimentally
show that, on all range queries, mzRTree outperforms other known structures used
for LC-MS data, even on those queries these structures are optimized for.
Besides, mzRTree is also more space efficient. As a result, mzRTree reduces data
analysis computational costs for very large profile datasets.
\end{abstract}


\end{frontmatter}

\section{Introduction}
Mass spectrometry-based proteomics~\cite{AebersoldM03} plays an ever-increasing
role in
different biological and medical fields, but, as an emerging field, it still
requires reliable tools for the storage, exchange and analysis of experimental
data. Over the last years, it has become available a wide range of technologies
which can generate a huge quantity of data potentially able to address relevant
questions, e.g., to identify proteins in a biological sample, to quantify their
concentration, to monitor post-translational modifications, to measure
individual protein turnover, to infer on interactions with other proteins,
transcripts, drugs or molecules. The technology is quickly advancing but,
without efficient bioinformatics tools, high-throughput proteomics data handling
and analysis are difficult and error-prone. Therefore, a major challenge facing
proteomic research is how to manage the overwhelming amount of data in order to
extract the desired information. This holds especially for high-throughput
quantitative proteomics, since it needs highly informative, high-resolution
profile data, in order to achieve reliable quantifications. Moreover, the data
hostage held by different instrument proprietary formats slows down the
evolution of proteomics, mainly because comparisons among different experiments,
or analytical methods, often become unfeasible.

In order to facilitate data exchange and management, the \textit{Human Proteome
Organization} (\textit{HUPO})~\cite{Orchard07} established the
\textit{Proteomics Standards
Initiative} (\textit{PSI}). HUPO-PSI released the \textit{Minimum Information
About a Proteomics Experiment} (\textit{MIAPE}) reporting guidelines
\cite{Taylor07} and
proposed \textit{mzData}~\cite{mzData}, which, as \textit{mzXML}
\cite{Pedrioli04}, is an
\textit{eXtensible Markup Language} (\textit{XML}) based data format, developed
to uniform data. Recently, merging the best features of each of these formats,
the HUPO introduced \textit{mzML} as a unique data format~\cite{Deutsch08}.
XML-based data
formats are characterized by intuitive language and a standardized structure. At
the state of art, the adoption of these formats is widespread among the
proteomics research groups, also thanks to the extensive support of instrument
and database searching vendors, and the availability of converters from
proprietary data formats. In spite of their success, the currently adopted
formats suffer from some limitations~\cite{Lin05}: the impossibility to store
raw data~\cite{Martens05}; the lack of information on the experimental design, necessary for regulatory
submission; the lack of scalability with respect to data size, a bottleneck for
the analysis of profile data. Above all, the 1-dimensional (1D) data indexing
provided by these formats considerably penalizes the analysis of datasets
embodying an inherent 2-dimensional (2D) indexing structure, such as
\textit{Liquid Chromatography-MS} (\textit{LC-MS}) ones.

LC-MS provides intensity data on a 2D $(t, m/z)$ domain, since LC separates
proteins along retention time dimension (\textit{temporal} index) based on their
chemical-physical properties, while MS separates proteins based on their mass
over charge ($m/z$ index) ratios. Minimizing the computational time to access
these huge datasets plays a key role in the progress of LC-MS data mining, and
can be of help also in a variety of other MS techniques, since MS experiments
usually have a “temporal” index related to the experimental time at which the MS
acquisition takes place (e.g., a \textit{scan} in mzXML). Therefore, MS data can
be accessed by means of either an $m/z$ range, or a temporal range, or a
combination of them, defining different range queries. On LC-MS data, these
accesses provide respectively \textit{chromatograms}, \textit{spectra}, and
\textit{peptide data}, whereas on generic MS data, they provide a set of
sub-spectra belonging to the specified range. An elevated number of range
queries are required during data analysis, thus optimizing them would
significantly improve computational performance.

Most research groups develop, often in a sub-optimal way, intermediate data
structures optimized for accesses on a privileged dimension, depending on the
downstream analysis. For instance, accredited software packages like
\textit{Maspectras}~\cite{Hartler07} or \textit{MapQuant}~\cite{Leptos06} make
use of the
method-specific intermediate data structures Chrom and OpenRaw, respectively:
the former is optimized for a chromatogram based access, the latter for a
spectra based access. In a recent work~\cite{Khan09} Khan et al. provide
evidence that
the use of a spatial indexing structure, namely the kd-tree, is suitable for
handling large LC-MS datasets and supporting the extraction of quantitative
measurements. The authors emphasize the effectiveness of the kd-tree for
performing analyses based on range queries but they do not compare explicitly
the range query performance of the kd-tree with that attainable by other known
data structures. Moreover, their experimental assessment is carried out only on
centroid datasets and does not consider profile data, which, as the literature
often remarks~\cite{Martens05}, are the most informative, especially for
quantitative
analysis, but also the most challenging to handle. For this reason the objective
of this work would rather be to develop a data structure to efficiently access
profile data.

Here, we present a novel data structure, called \textit{mzRTree}, for
high-throughput LC-MS profile datasets, which combines a hybrid sparse/dense
matrix representation of the data and a scalable index based on the R-tree
\cite{Guttman84}.
We show experimentally that mzRTree supports efficiently both 1D and 2D data
accesses. In particular, mzRTree significantly outperforms Chrom and OpenRaw on
small and large peptide range queries, yielding in some cases orders of
magnitude improvements. Furthermore, it still ensures best performance on the
accesses for which the other data structures are optimized, i.e., chromatograms
for Chrom and spectra for OpenRaw. The experiments also provide evidence that
mzRTree is more space efficient than Chrom and OpenRaw, and exhibits good
scalability on increasing dataset densities.

\section{The mzRTree Data Structure}
Let us conceptually view an LC-MS dataset $D$ as a matrix, where the rows are
indexed by retention times, the columns by $m/z$ values, and the entries are
intensity values. A generic entry is denoted as $(rt, mz; I)$, where $rt$ and
$mz$ are the row and column indices, and $I$ is the intensity value.

We store $D$ using a hybrid sparse/dense matrix representation, as follows.
First, we evenly subdivide the matrix into $K$ \textit{strips} of consecutive
rows, where $K$ is a user defined parameter. Then, each strip is in turn
partitioned into a number of bounding boxes (BBs), each corresponding to a
distinct range of $m/z$ values. In our implementation, each BB corresponds to
approximately $5$ Da, and $K$ is set in such a way to ensure that each strip
fits in the main memory (RAM). A BB is characterized by four coordinates,
namely: top-rt (resp., bottom-rt), which is the smallest (resp., largest)
retention time of the BB’s nonzero intensity entries; and left-mz (resp.,
right-mz), which is the smallest (resp., largest) $m/z$ value of the BB’s
nonzero intensity entries. The BBs of a strip are stored consecutively in a
file, and each strip is saved in a distinct file so that it can be efficiently
loaded in the main memory during a range query. If half or more of the entries
in a BB have nonzero intensity, then the BB is stored in the file using a dense
matrix representation. Otherwise, a sparse representation is used storing the
nonzero intensity entries in row-major order, indicating for each such entry the
column ($m/z$ value) and the intensity, and separating successive rows through
special characters. In this fashion, each BB occupies a space proportional to
the number of nonzero intensity entries it contains.

A \textit{range query operation} on $D$ takes as input two retention times
$rt_1$, $rt_2$, and two $m/z$ values, $mz_1$, $mz_2$, and returns all entries
$(rt, mz; I)$ in $D$ such that $rt_1 < rt \leq  rt_2$ and $mz_1 < mz \leq mz_2$.
Accesses to chromatograms, spectra or peptide data can be easily expressed
through range queries. In order to support efficient range query operations, we
use an index implemented through a tree structure based on the
\textit{R-tree}~\cite{Guttman84,Vitter01}, which is a well known spatial data
structure for managing geometric data.

Let $d$ and $f$ be two integer parameters, and let $G$ be the set of nonempty
BBs (i.e., BBs which contain at least one nonzero-intensity entry). Denote by
$W$ the cardinality of $G$. Our index consists of a balanced search tree whose
leaves are associated with disjoint subsets of $G$ forming a partition of $G$.
The number of children of each internal node is proportional to d (the root, if
internal, may have a smaller number of children) and each leaf is associated
with a subset of size proportional to $f$ of BBs in $G$ (the root, if a leaf,
may have less than $f$ BBs). Each internal node of the tree is associated to the
smallest submatrix of D which contains all BBs associated with its descendant
leaves.

The execution of a range query requires to traverse all root-leaf paths ending
in leaves associated with BBs that intersect the rectangle defined by the query,
and to return all entries of interest. The complexity of a range query depends
on the height of the tree, hence on the parameters $d$ and $f$, and on the
mapping of the BBs to the leaves. As for the choice of the partition parameters
$d$ and $f$, when dealing with massive datasets, which must be kept in secondary
memory, it is convenient to impose that each node of the tree (except, possibly,
the root) occupies a constant fraction of the minimal unit that can be
transferred between the secondary memory and the RAM. Instead, for what concerns
the mapping of the BBs to the leaves, several heuristics have been proposed in
the literature (see~\cite{Vitter01} for relevant references).

In our implementation, we set $d=6$ and $f=200$, and the actual structure of the
tree is recursively defined as follows, based on ideas in~\cite{Guttman84}. If
$W\leq f$,
the tree consists of a single leaf associated with the set $G$; otherwise, $G$
is partitioned into six groups, $G_i$, for $1 \leq  i \leq  6$, as follows.
$G_1$ contains the $\lceil W/6\rceil$ BBs with smallest top-rt; $G_2$ contains
the $\lceil W/6\rceil$ BBs in $G-G_1$ with smallest left-mz; $G_3$ contains
the $\lceil W/6\rceil$ BBs in $G- G1- G2$ with largest bottom-rt; $G_4$
contains the $\lceil W/6\rceil$ BBs in $G- G1- G2- G3$ with largest right-mz;
$G_5$ contains the $\lceil W/6\rceil$ BBs in $G- G1- G2- G3- G4$ with smallest
left-mz; and $G_6$ contains the remaining BBs. The six groups are associated
with the subtrees rooted at the children of the root, which are recursively
organized in a similar fashion. Each leaf is thus associated with up to $f=200$
BBs, and it stores, for each of its BBs, the four coordinates (top-rt,
bottom-rt, left-mz, right-mz) and a pointer to the file where the BB is
stored together with the relative offset within the file. It can be easily shown
that the height of the tree is proportional to $\log_6 (W/200)$.

We call \textit{mzRTree} the whole data structure, which includes the actual
data (i.e.,
the bounding boxes) stored in the files, and the tree index described above. We
developed a Java implementation of mzRTree, which includes a method to build an
mzRTree starting from an input dataset provided in mzXML format
\cite{Pedrioli04}, and a
method to perform a generic range query\footnote{The Java code implementing
mzRTree is available for download at \url{http://www.dei.unipd.it/mzrtree}.}.

\section{Experimental Results and Discussion}
In this section, we evaluate and discuss mzRTree performance compared to Chrom
and OpenRaw, which are two existing data structures used by Maspectras
\cite{Hartler07} and
MapQuant~\cite{Leptos06} software packages and optimized for chromatograms and
spectra
based accesses, respectively. Specifically, we focused our analysis on the time
required for a range query, the time required for building up the data
structure, and the required hard disk space. Furthermore, we verified mzRTree
scalability for what concerns range query times using datasets of increasing
density, where the density of a dataset is defined as the ratio of the number of
retention time and $m/z$ value pairs associated with nonzero intensities to the
overall number of retention time and $m/z$ value pairs.

We compared mzRTree, Chrom and OpenRaw on seven LC-MS datasets, named EXP1,
EXP2, ART1, ART2, ART3, ART4 and ART5, which are described below. The
\textit{EXP1} dataset consists of real profile data from a controlled mixture of
ICPL-labeled proteins acquired in enhanced profile mode for survey scans to gain
higher mass accuracy using a Finnigan LTQ linear ITMS (Thermo Electron) equipped
with HPLC-NSI source. The \textit{EXP2} is a real profile dataset acquired with
a Waters ESI TOF Microchannel plate LCT Premier available on the PeptideAtlas
public database. The \textit{ART1}, \textit{ART2} and \textit{ART3} datasets
have been generated by the LC-MS simulator LC-MSsim~\cite{TrieglaffPGKR08} using
as input some
peptide sequences from bovine serum albumin (UniprotKB: P02769), human
apotransferrin (UniprotKB: P02787) and rabbit phosphorylase b (UniprotKB:
P00489). Finally, the \textit{ART4} and \textit{ART5} datasets have been
generated artificially by the following procedure: for each dataset, the user
specifies some input parameters, namely, the number of spectra (i.e., the total
number of retention times), the $m/z$ range and resolution, and the density d;
then, each spectrum is populated by assigning nonzero intensity values to
positions corresponding to $m/z$ values drawn from a uniform distribution until
the density of the spectrum is $d$; clearly, if each spectrum has density $d$,
then the final dataset will have density $d$. ART4 and ART5 are useful to
evaluate the scalability of our data structure although they are not meaningful
from a biological standpoint.

The characteristics of the aforementioned datasets are summarized in
Table~\ref{table1}. Notice that the resolution shown in Table~\ref{table1} is
not the original data resolution (i.e., the instrumental resolution) but it is a
suitable resolution, not smaller than the original one, which has been adopted
in our data representation for uniformity with the other data representations
used for comparison in the experiments. In particular, the Chrom files we used,
adopted a $0.001$ Da resolution: this resolution is higher than the maximum
resolution achievable by the instruments used to acquire the experimental data.
Therefore, our choice is conservative in the sense that it does not require any
binning and, consequently, does not cause any loss of information.

\begin{table}
 {\small 
\newcommand{\mc}[3]{\multicolumn{#1}{#2}{#3}}
\begin{center}
\begin{tabular}{p{0.09\textwidth}|p{0.09\textwidth}|p{0.09\textwidth}|p{0.09\textwidth}|p{0.09\textwidth}|p{0.09\textwidth}|p{0.09\textwidth}|p{0.09\textwidth}|}\cline{2-8}
 & \textbf{EXP1} & \textbf{EXP2} & \textbf{ART1} & \textbf{ART2} & \textbf{ART3} & \textbf{ART4} & \textbf{ART5}\\\hline
\mc{1}{|l|}{\textbf{type}} & real & real & artificial & artificial & artificial & artificial & artificial\\\hline
\mc{1}{|l|}{\textbf{$\mathbf{m/z}$ range}} & $400-1800$ & $400-1600$ & $400-1800$ & $400-1800$ & $400-1800$ & $400-1800$ & $400-1800$\\\hline
\mc{1}{|l|}{\textbf{spectra number}} & $2130$ & $6596$ & $2400$ & $2400$ & $2400$ & $2130$ & $2130$\\\hline
\mc{1}{|l|}{\textbf{resolution}} & $0.001$ & $0.001$ & $0.001$ & $0.001$ & $0.001$ & $0.001$ & $0.001$\\\hline
\mc{1}{|l|}{\textbf{density}} & $2.50\%$ & $6.27\%$ & $2.56\%$ & $4.05\%$ & $8.27\%$ & $16.00\%$ & $28.00\%$\\\hline
\mc{1}{|l|}{\textbf{mzXML file size}} & $769 MB$ & $5 GB$ & $657 MB$ & $1 GB$ & $2.8 GB$ & $4.8 GB$ & $8.3 GB$\\\hline
\end{tabular}
\end{center}
}%
\caption{\textit{Datasets' Features.} Notice that the spectra number is referred to the total number of MS1 spectra and resolution is not the instrument resolution, as explained in the text.\label{table1}}
\end{table}

\begin{figure}
 \centering
 \includegraphics[width=.8\textwidth]{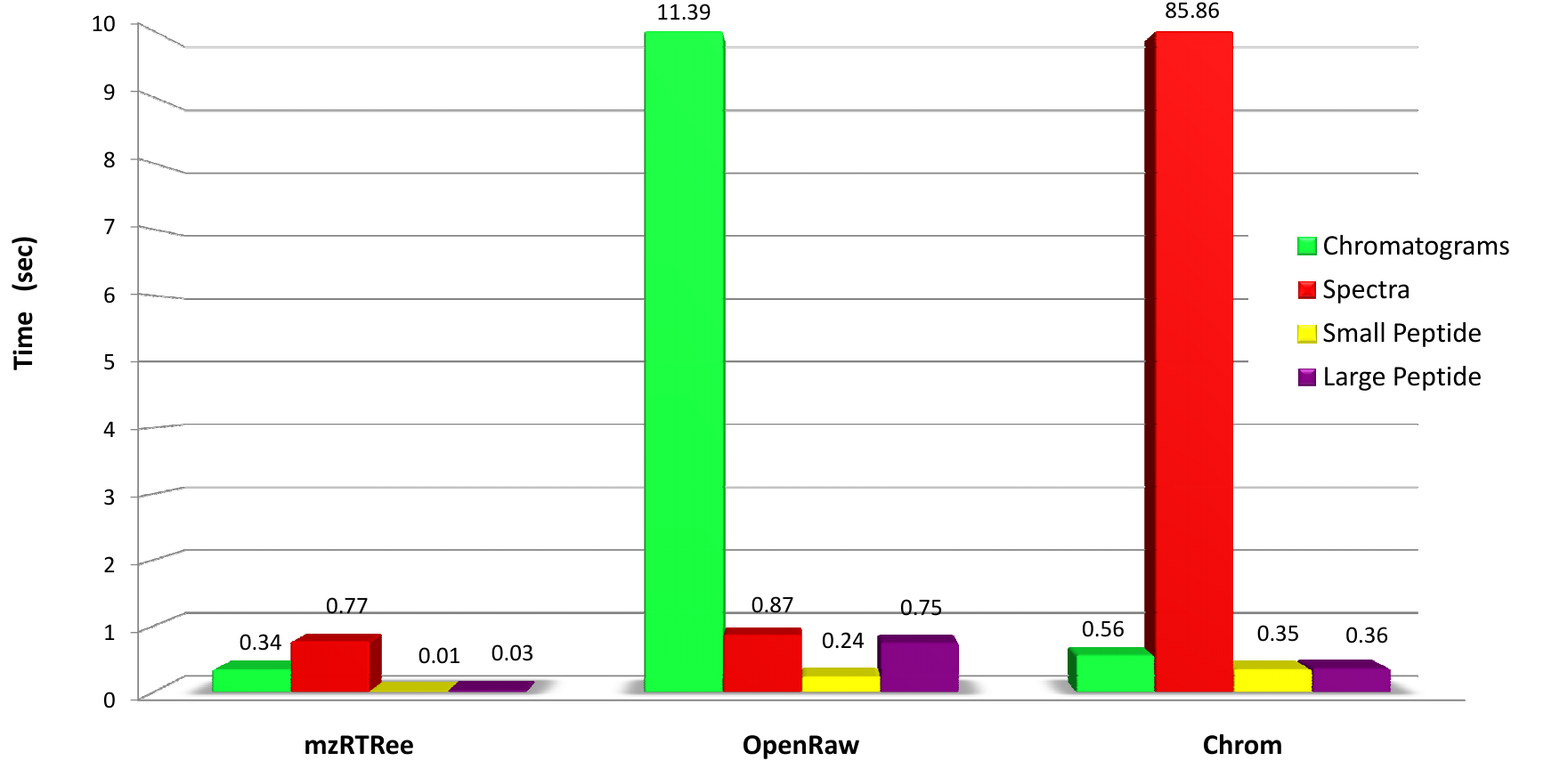}
 \caption{\label{fig1}\textit{Real Data Access Times On EXP1.}  Comparison on EXP1 dataset among mzRTree, OpenRaw and Chrom on random chromatograms, spectra and small/large peptide range queries spanning the whole dataset as regards access times. Every colored bar refers to a different kind of range query. mzRTree reaches best performance on all kind of range queries, outperforming Chrom and OpenRaw.}
\end{figure}

\begin{figure}
 \centering
 \includegraphics[width=.8\textwidth]{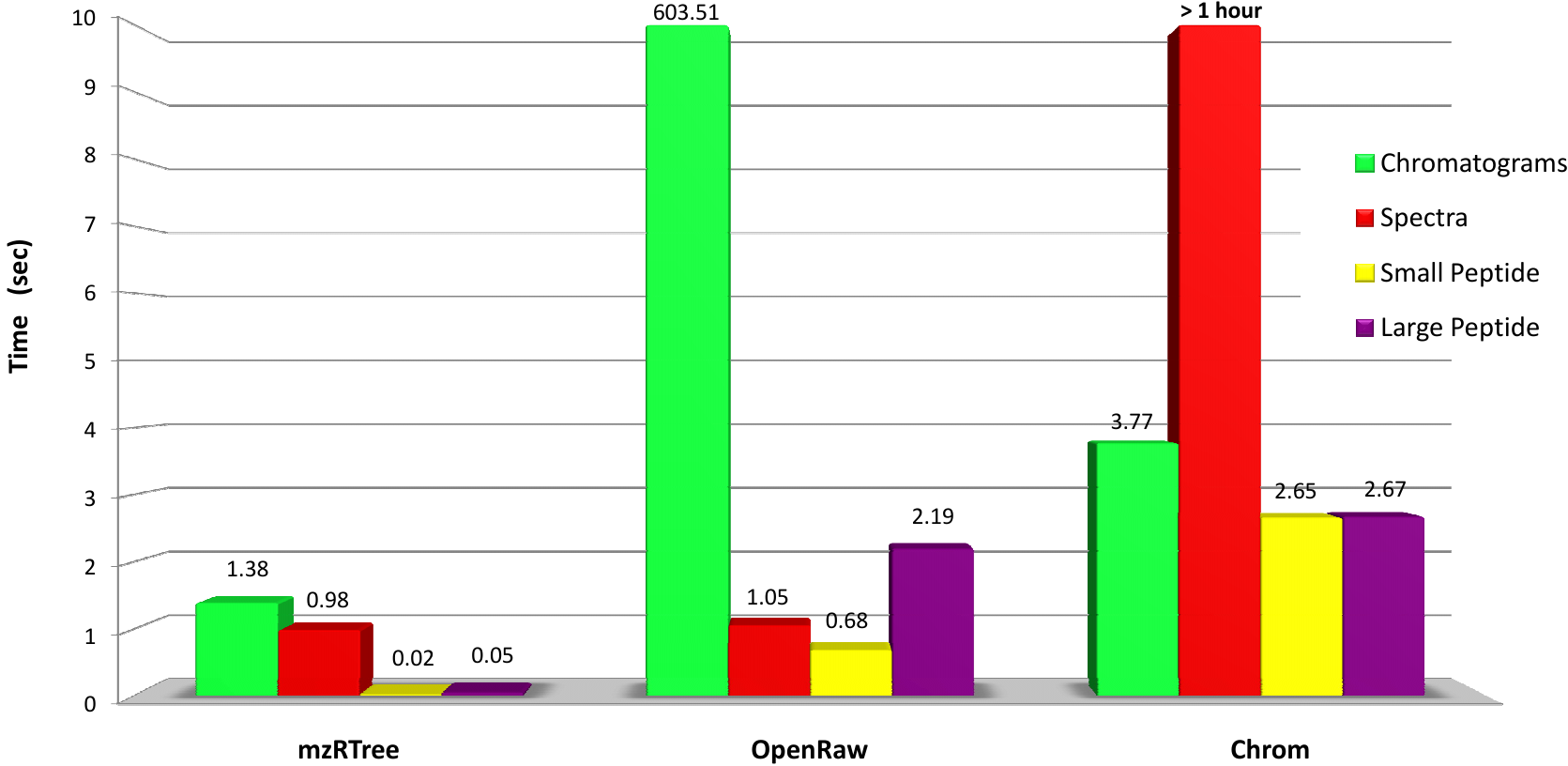}
 \caption{\label{fig2}\textit{Real Data Access Times On EXP2.} Comparison on EXP2 dataset among mzRTree, OpenRaw and Chrom on random chromatograms, spectra and small/large peptide range queries spanning the whole dataset as regards access times. Every colored bar refers to a different kind of range query. Notice how mzRTree still reaches best performance, outperforming Chrom and OpenRaw, also on this higher density and size dataset.}
\end{figure}

\begin{figure}
 \centering
 \includegraphics[width=.8\textwidth]{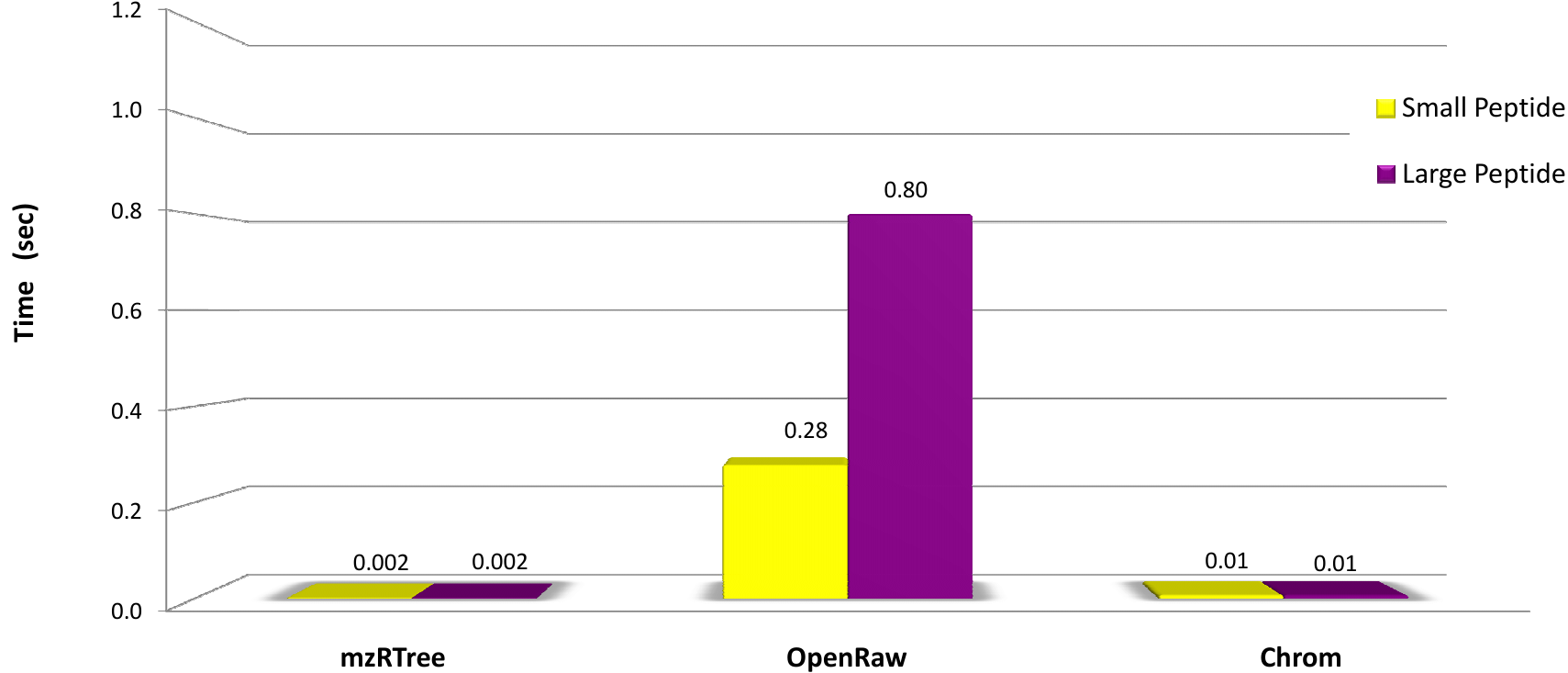}
 \caption{\label{fig3}\textit{Real Data Access Times For Mascot Identified Peptides On EXP 1.} Comparison on EXP1 dataset among mzRTree, OpenRaw and Chrom on small/large peptide range queries related to Mascot identified peptides as regards access times: mzRTree is one order of magnitude faster than Chrom and two orders of magnitude faster than OpenRaw.}
\end{figure}

We compared mzRTree, Chrom and OpenRaw on four kinds of range queries: a
rectangle covering all the retention times and a $5$ Da range in the $m/z$
dimension (\textit{chromatograms}); a rectangle covering the entire $m/z$
dimension and $20$ retention times (\textit{spectra}); a rectangle of $5$ Da and
$60$ retention times (\textit{small peptide}); a rectangle of $5$ Da and $200$
retention times (\textit{large peptide}). We estimated the performance for each
kind of range query summing the access times required to perform ten range
queries spanning the whole dataset in order to avoid any local advantage. More
precisely, we evaluated separately the time required for loading the internal
variables used by each data structure every time it is invoked (\textit{load
time}) and the time actually needed to perform only the range query
(\textit{access time}). To reduce random variability, we computed both access
and load times averaging over ten experimental repetitions. It is worth to
notice that a spectra range query is more time consuming than a chromatograms
range query, since the number of distinct $m/z$ values is typically much bigger
than the number of retention times. Results on access times for the EXP1 and
EXP2 datasets are shown in Figures~\ref{fig1} and~\ref{fig2}, respectively. As
the figures clearly show, mzRTree achieves the best performance on all kinds of
range queries for both the smaller size and density dataset EXP1 and the larger
size and density dataset EXP2. Furthermore, Figure~\ref{fig3} illustrates the
access times for ten peptides in EXP1 using small and large peptide range
queries, whose bounds refer to peptides actually identified by the Mascot search
engine. mzRTree significantly outperforms Chrom and OpenRaw on small and large
peptide range queries, and still ensures best performance on the accesses for
which the other data structures are optimized, i.e., chromatograms for Chrom and
spectra for OpenRaw.

\begin{figure}
 \centering
 \includegraphics[width=.8\textwidth]{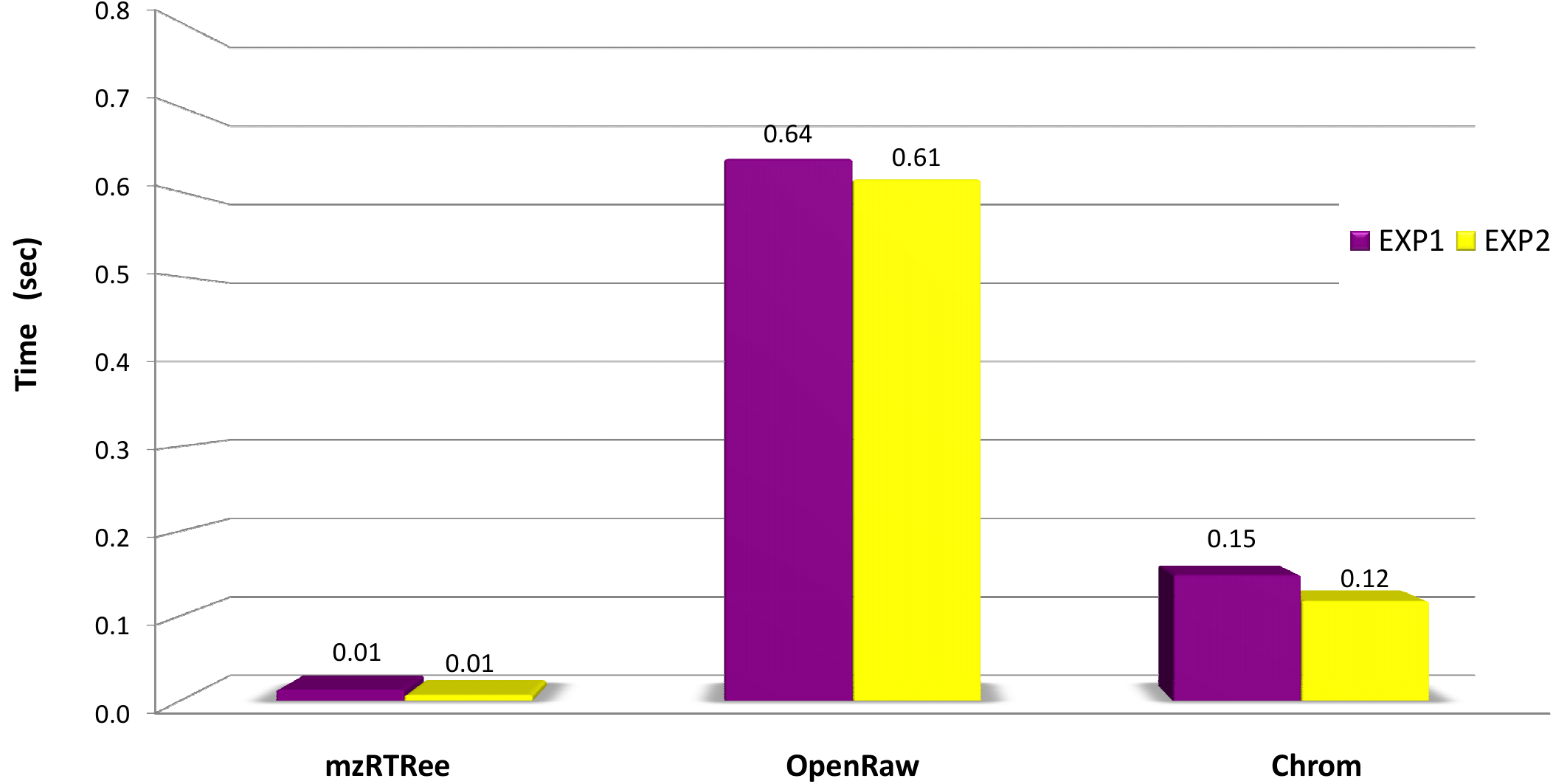}
 \caption{\label{fig4}\textit{Load Times For EXP1 And EXP2.} Comparison on EXP1 and EXP2 datasets among mzRTree, OpenRaw and Chrom on load times: mzRTree is one order of magnitude faster than Chrom and OpenRaw.}
\end{figure}

The load time required by the three data structures is shown in
Figure~\ref{fig4} for EXP1 and EXP2 datasets: we note that the load time is
mainly independent of dataset features, and mzRTree still achieves the best
performance. Since loading is required every time the data structures are
invoked, it is convenient to perform many consecutive range queries in order to
amortize its cost: the higher the load time, the more the range queries needed
to amortize it.

\begin{figure}
 \centering
 \includegraphics[width=.8\textwidth]{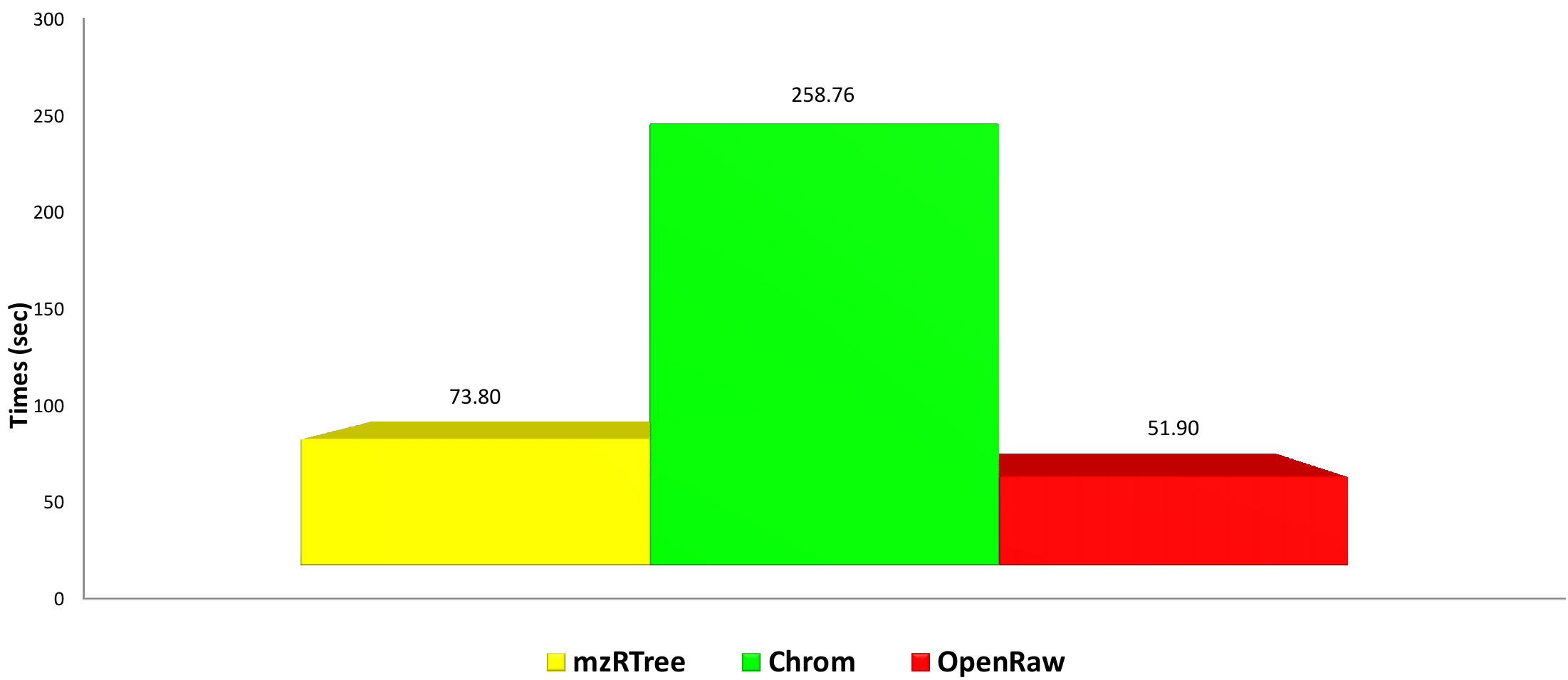}
 \caption{\label{fig5}\textit{Data Structures' Creation Time For EXP1.} Comparison of mzRTree, Chrom and OpenRaw as regards data structures’ creation time for EXP1 dataset. }
\end{figure}

Even if the data structure creation takes place just once, we also estimated the
creation time for mzRTree, Chrom and OpenRaw on EXP1. Notice that while mzRTree
and Chrom creation starts from the mzXML file, the OpenRaw creation starts from
the .RAW file, requiring the instrument vendor’s software to be licensed and
installed on the computer. We chose EXP1 because its size is small enough to fit
in RAM, thus all three data structures evenly work at their best condition. As
shown in Figure~\ref{fig5}, mzRTree features an efficient creation time, even if
OpenRaw reaches the best performance. However, notice that OpenRaw is advantaged
since it starts from binary data instead of Base64 encoded data.

\begin{figure}
 \centering
 \includegraphics[width=.8\textwidth]{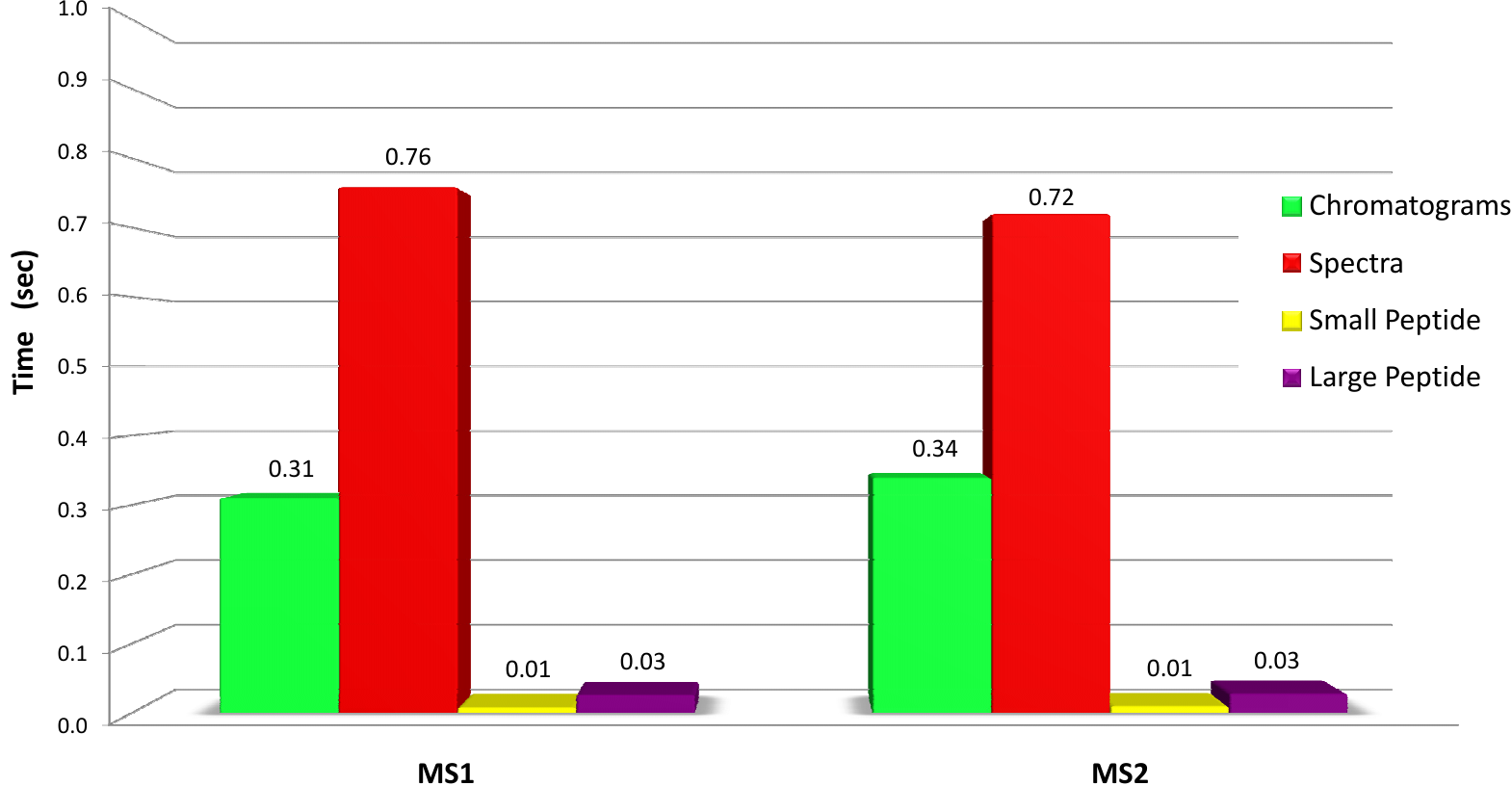}
 \caption{\label{fig6}\textit{Real Data Access Times On Ms$^1$ \& Ms$^2$ For EXP1.} Comparison of mzRTree access times on MS$^1$ and MS$^2$ levels for EXP1 dataset. The performance of  mzRTree is independent of the MS level.}
\end{figure}

In Table~\ref{table2} we provide the comparison of the space reduction using
mzRTree, Chrom and OpenRaw compared to the mzXML hard disk space, which we chose
as reference. mzRTree requires the smallest amount of space, hence it allows for
cheaper storage and easier sharing of proteomics datasets. Besides, mzRTree
storage requires at least $30\%$ less hard disk space than XML based data
formats, since mzRTree stores binary data instead of Base64 encoded data: it is
a considerable amount of space saved, when taking into account RAID systems and
backup systems. Observe that, since, for the sake of simplicity, we are ignoring
MS level-two spectra, the space savings for the first two datasets are notably
larger than 30\%; however, this is not the case of the third dataset, which
consists only of level-one spectra. Anyway, mzRTree can efficiently handle also
tandem data; the user only needs to create the data structure for every MS/MS
level of interest. Figure~\ref{fig6} shows that mzRTree provides efficient
access times on tandem MS data for all kind of range queries, attaining for MS
level $2$ data the same performance as for MS level $1$ data.

\begin{table}
\small
\begin{center}
\begin{tabular}{|l|l|l|l|}\hline
\textbf{mzXML} & \textbf{EXP1} & \textbf{EXP2} & \textbf{ART4}\\\hline
\textbf{mzRTree} & $3.71\%$ & $46.00\%$ & $25.00\%$\\\hline
\textbf{Chrom} & $37.84\%$ & $28.00\%$ & $-$\\\hline
\textbf{OpenRaw} & $7.31\%$ & $18.00\%$ & $-10.42\%$\\\hline
\end{tabular}
\end{center}
\caption{\textit{Hard Disk Space Savings.} Space reduction referred to the original mzXML file size, chosen as reference. mzRTree allows for a more efficient hard disk space-saving storage.\label{table2}}
\end{table}

\begin{figure}
 \centering
 \includegraphics[width=.8\textwidth]{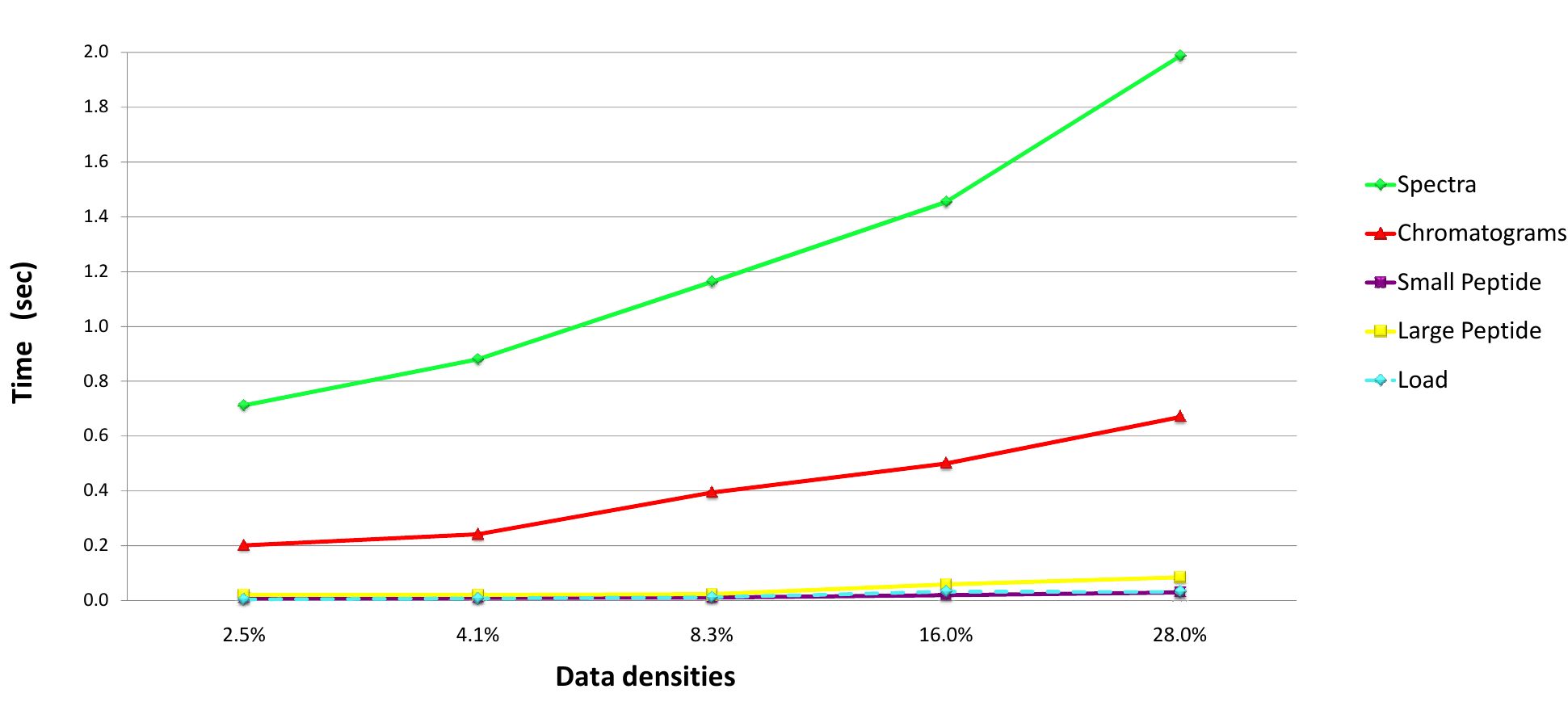}
 \caption{\label{fig7} \textit{Access \& Load Times On Simulated Data.} Evaluation of mzRTree scalability on increasing dataset densities as regards the load time and access times on different kind of range queries. }
\end{figure}

To test mzRTree scalability on increasing dataset densities and sizes we
performed different range queries on the artificial datasets ART1, ART2, ART3,
ART4 and ART5. Results are illustrated in Figure~\ref{fig7}, which shows that
mzRTree is fairly scalable as regards access and load time: as data density
increases by a factor $10$, the access time increases only by a factor $3$ in
the worst case, while the load time is almost constant.

These results and the inherent scalability of the underlying R-tree structure
suggest that mzRTree is suitable for high density/large size proteomics data,
such as profile data, considered as the most informative and hence the most
suitable to tackle quantification aims~\cite{Lin05,Martens05}. At present, profile data size
reaches some GB, but it is expected to further increase, as far as instrument
accuracy and resolution increase: even a narrow range of $m/z$ values can be
challenging to manage when analyzing these data. Thus, the adoption of mzRTree
for data storage could make profile data accessible for analysis purposes: it
prevents out-of-memory errors, often occurring with huge profile proteomics
datasets, and reduces the need for (and the costs of) extraordinary
computational infrastructures and their management. Actually, profile data are
often the only data source rich enough to carry on a meaningful analysis, e.g.,
in quantitative proteomics based on stable isotope labelling. However, costs
involved with profile data handling often outweigh their benefits. mzRTree could
revert this relationship.

\section{Conclusions}
In this paper we proposed mzRTree, a scalable and memory efficient spatial
structure for storing and accessing LC-MS data, which features efficient
construction time and faster range query performance, compared to other known
and widely used data structures.

Several research questions remain open. The efficiency of mzRTree depends on
several design choices, including the degree of the internal nodes and the way
the bounding boxes are mapped to the leaves of the tree. The design space for
mzRTree should be fully explored in order to identify the best choices.
Moreover, when dealing with huge raw datasets mzRTree may not fit in RAM. In
that case, the tree must reside on hard disk and the size of the internal nodes
should be adapted to match the minimum block size used in disk-RAM data
movements. Other solutions based on indexing structures alternative to the
R-tree employed by mzRTree (e.g., those surveyed in~\cite{Vitter01}, including
the kd-tree
used in~\cite{Khan09}) should be considered and compared to mzRTree. Finally, it
is
interesting and potentially useful to investigate effective ways to further
integrate all additional information needed for regulatory submission into
mzRTree.

\section{Acknowledgements}
A special thank to Juergen Hartler for providing us with the ICPL datasets and
with code for Chrom, to Piero De Gol for his informatics support. We would like
to thank also
the anonymous referees for their useful comments, which helped us improving the
presentation. This study was supported by funding of both the CARIPARO 2008/2010
``Systems biology approaches to infer gene regulation from gene and protein time
series data'' project and the University of Padova Strategic Project STPD08JA32
``Algorithms and Architectures for Computational Science and Engineering''.

\end{document}